\documentclass{PoS}
\usepackage{cleveref}
\usepackage{wrapfig}
\usepackage{placeins}
\usepackage{wasysym}
\usepackage{lineno}

\title{First results of the deployment of a SoLid detector module at the SCK$\bullet$CEN BR2 reactor}

\ShortTitle{First results of the deployment of a SoLid detector module at the SCK$\bullet$CEN BR2 reactor}

\author{\speaker{Nick Ryder}%
         \thanks{On behalf of The SoLid Collaboration.}\\
        University of Oxford (GB)\\
        E-mail: \email{nick.ryder@physics.ox.ac.uk}}

\abstract{
The SoLid experiment aims to resolve the reactor neutrino anomaly by searching for electron-to-sterile anti-neutrino oscillations.
The search will be performed between 5.5 and 10 m from the highly enriched uranium core of the BR2 reactor at SCK$\bullet$CEN.
The experiment utilises a novel approach to anti-neutrino detection based on a highly segmented, composite scintillator detector design.
High experimental sensitivity can be achieved using a combination of high neutron-gamma discrimination using $^6$LiF:ZnS(Ag) and precise localisation of the inverse beta decay products.
This compact detector system requires limited passive shielding as it relies on spacial topology to determine the different classes of backgrounds.
The first full scale, 288 kg, detector module was deployed at the BR2 reactor in November 2014.
A phased three tonne experimental deployment will begin in the second half of 2016, allowing a precise search for oscillations that will resolve the reactor anomaly using a three tonne detector running for three years.
In this talk the novel detector design is explained and initial detector performance results from the module level deployment are presented along with an estimation of the physics reach of the next phase.
}

\FullConference{The European Physical Society Conference on High Energy Physics\\
                 22-29 July 2015\\
                 Vienna, Austria}

\begin{document}

\crefname{figure}{figure}{figures}

\section{Introduction}


The SoLid experiment aims to resolve the reactor neutrino anomaly.
The reactor anomaly arose when the reactor anti-neutrino flux was recalculated for the latest generation of short baseline reactor anti-neutrino experiments aimed at measuring $\theta_{13}$~\cite{reactorcalculations}.
Using the updated calculations of the flux and comparing them to measurements performed in the 1980s and 1990s, a $2.5 \sigma$ deficit in the measured flux became apparent~\cite{reactoranomaly}.

Due to the neutrino's non-zero mass it is possible that the deficit is can be explained by the electron anti-neutrino emitted by the reactor oscillating into another flavour state.
This would have resulted in a measured deficit since the experiments were only sensitive to electron flavour anti-neutrinos, via the Inverse Beta Decay (IBD) reaction.
The deficit could be explained by an oscillation of approximately 8\% of the 1.8 - 10 MeV electron anti-neutrinos within 100 m from their production in the reactor cores.
Oscillations to muon or tau neutrinos cannot explain such a deficit, and therefore a new flavour of neutrino would be required for oscillations to explain the reactor anomaly.
The new neutrino is called `sterile' since it is known that only 3 flavours interact with the $Z$ boson \cite{Zneutrinos}.

A similar deficit was measured when intense radioactive sources were used to calibrate the SAGE and GALLEX solar neutrino experiments~\cite{sage, gallex}.
An oscillation from electron anti-neutrinos to a sterile neutrino flavour can be used to explain both the reactor and gallium anomalies, with the best fit for possible oscillation parameters being $\sin^2(2\theta_s) \approx 0.1$ and $\Delta m^2_s \approx 1 eV^2$~\cite{reactoranomaly}.

In the reactor and gallium anomalies the measured integral flux at a given distance was compared to a calculated expectation value.
To resolve these anomalies and determine whether they are caused by oscillations it is necessary to measure the anti-neutrino flux as a function of both energy and distance.
A direct search can then be made for oscillations, without relying on any calculations of the source's anti-neutrino flux or its energy spectrum.

\begin{figure}
    \centering
    \includegraphics[width=0.75\textwidth]{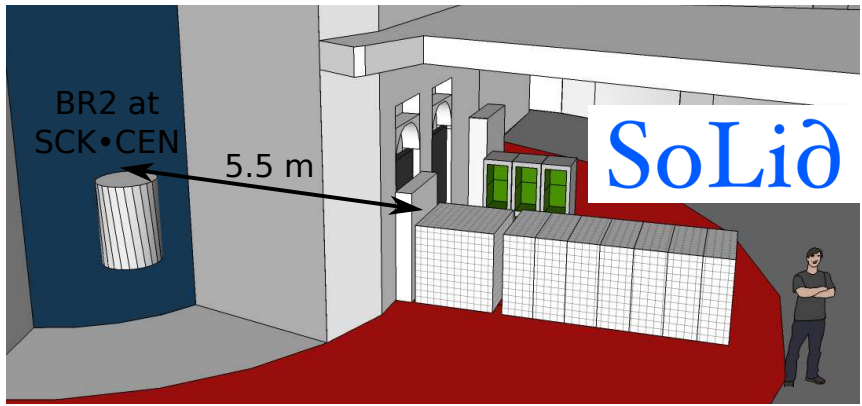}
    \caption{Diagram of the SoLid experiment deployed 5.5 m from the BR2 reactor core. The Detector is split into multiple modules of two different technologies, explained in the text.}
    \label{fig:solidatBR2}
\end{figure}

The environment close to a nuclear reactor raises a number of experimental challenges.
There are high rates of background events due to the low over burden and close proximity to the reactor.
Cosmic ray muons can cause fast spallation neutrons.
These can mimic an IBD event when a proton recoils from the neutron, producing a $e^+$-like signal, which is followed by the thermal neutron being captured.
Cosmic rays also create cosmogenic isotopes in the detector.
Isotopes such as $^8$He or $^9$Li lead to $\beta^- + n$ decays, where the $\beta^-$ can fake the positron signal and the thermal neutron is also detected.
There are also backgrounds due to accidental time coincidences of randomly distributed background $\gamma$ rays with environmental neutrons.
This background is increased by additional neutrons and $\gamma$ rays that can be emitted when the reactor is running.
As well as the high rate of background events, there are also constraints on the materials that can safely be used (in particular the combination of flammable liquids and high voltages is discouraged), the size of the detector, the accessibility and data handling infrastructure due to the deployment as close as possible to the reactor core.

The SoLid experiment, shown in \cref{fig:solidatBR2}, will search for such an oscillation using a three tonne, highly segmented, composite scintillator anti-neutrino detector deployed between 5.5 and 10 m from the centre of the BR2 reactor's highly enriched uranium core.
The BR2 reactor is designed for neutron irradiations for material testing, medical isotope production and doping of semiconductors.
The reactor has a unique twisted rod design to provide a compact core (with fission $\diameter_{rms} < 0.5$ m) whilst providing access for deploying material within the high neutron flux.
It usually runs c. 70 MW cycles lasting 3-4 weeks, running approximately 50\% of the time.
The BR2 reactor provides a relatively low background compared to other facilities since it is not being used as a beam source for other experiments and so has complete shielding around the core.

The approach to overcoming the experimental challenges was to develop a novel detector technology that has a robust neutron identification with a rich data set that can be used to efficiently identify IBD candidate events while minimising the backgrounds.
The novel detection concept is explained in \cref{sec:concept}.
This approach allows a detector with minimal passive shielding to perform the experiment, but required the production and deployment of prototype detectors to demonstrate and understand the new technology.
A 288 kg full scale detector module was deployed at the BR2 reactor in November 2014, which took reactor on, reactor off and calibration source data between February and September 2015.
The 288 kg module and its deployment are detailed in \cref{sec:deployment}.
The analysis of the data provided by this module is ongoing, with initial results given in \cref{sec:initialresults}.

Starting in 2016 the SoLid collaboration will performed a phased experiment that will provide high sensitivity to oscillations within 10 m of a nuclear reactor.
The plans for this experiment and the estimated sensitivity are presented in \cref{sec:futureplans}.

\section{Detector concept}
\label{sec:concept}

\begin{figure}[h]
    \centering
    \includegraphics[width=0.8\textwidth]{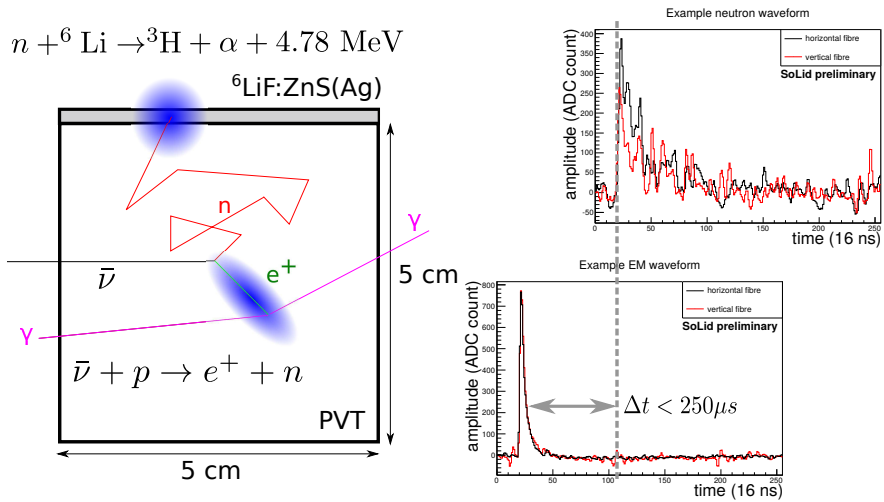}
    \caption{Comparison of scintillation signals from neutron capture (upper) in the $^6$LiF:ZnS(Ag) and electromagnetic ($e^+/\gamma/\mu$) electromagnetic ($e^+/\gamma/\mu$) signals in the PVT scintillator (lower).}
    \label{fig:pulsecomparison}
\end{figure}

The SoLid detector identifies the inverse beta decay products when an electron anti-neutrino interacts with a proton in the mass of the detector.
A composite scintillator design is used to unambiguously discriminate between captured neutrons and electromagnetic energy deposits, as illustrated in \cref{fig:pulsecomparison}.
The bulk of the detector (the neutrino target) is constructed from 5 cm cubes of polyvinyl toluene (PVT) scintillator.
The positron emitted from the IBD event causes the PVT to scintillate and then annihilates with an electron in the detector, emitting a pair of 511 keV $\gamma$ rays.
An example scintillation light pulse from a particle scintillating in the PVT, shown in the lower waveform in \cref{fig:pulsecomparison}.
The annihilation $\gamma$ rays deposit a small fraction of their energy in the interaction cube, and the rest in nearby PVT cubes.
The energy depositions in nearby cubes can be used as a positron tag, reducing the impact of background events.
In the IBD event the positron energy is directly related to the anti-neutrino energy, after correcting for the energy required to convert a proton into a more massive neutron.
In the SoLid detector the energy deposited in the interaction cube can be used as a good estimate of the positron energy with only a small correction for the additional energy deposited by the annihilation $\gamma$ rays.

Each PVT cube has a neutron sensitive $^6$LiF:ZnS(Ag) layer placed on one face.
The neutron from the IBD thermalises in the PVT and has approximately 50\% probability to be captured by a $^6$Li atom in the neutron sensitive layer, resulting in the interaction: $n + ^6\mbox{Li}\rightarrow \alpha + ^3\mbox{H} + 4.78 \mbox{ MeV}$.
The $\alpha$ and $^3$H particles deposit most of their energy in the ZnS(Ag) inorganic scintillator which is mixed with the $^6$LiF.
Due to the high energy density they excite states where the scintillation photons are slowly emitted, up to microseconds after the neutron capture, as shown in the upper waveform in \cref{fig:pulsecomparison}.
This scintillation signal can easily be distinguished from the single light pulse from PVT scintillation.
The differing time signatures for the light pulses can therefore be used to identify whether a given scintillation signal was produced in the ZnS (characteristic of a neutron capture) or in the PVT ($e^+/\gamma/\mu$-like).

In the 2 tonnes of the detector that will initially be deployed in 2016, the PVT cubes are wrapped in reflective Tyvek with the $^6$LiF:ZnS(Ag) layer inside the wrapping.
There are two $5\times5$ mm$^2$ grooves cut into each cube into which $3\times3$ mm$^2$ square wavelength shifting (WLS) fibres are placed.
The scintillation from both the PVT and the ZnS(Ag) is captured in the WLS fibres which are inserted in a 2D (vertical and horizontal) array, coupling multiple cubes to each fibre.
The light is detected by $3\times3$ mm$^2$ silicon photomultipliers (SiPM), optically coupled to one end of each fibre.
The opposite end of each fibre is coupled to a mirror.
In the third tonne, to be deployed in 2017, the PVT is doped with wavelength shifter and no optical isolation or optical fibres are used to separate the cubes.
The light is transported to the edge of the fibre by total internal reflection from the surfaces of the cubes and detected by 5 cm photomultiplier tubes (PMTs).
This increases the light collection, providing a better energy resolution.

In both light collection schemes the original location of the scintillation light can be determined to the cube level within each plane.
As such the neutron signal and the positron candidate signal can be located to within 5 cm.
In conventional neutrino detectors only the time difference between the positron candidate and the neutron capture can be used to select IBD events.
However the high level of segmentation in the SoLid detector allows the IBD candidate events to be imaged spatially as well.
The additional information provided by the segmentation is a powerful tool in selecting true IBD events from backgrounds.
For example the distance between the positron signal and neutron signal can be limited to the diffusion distance of the IBD neutrons within the detector, which is less than 15 cm.
The scintillation due to Compton scattering of the annihilation $\gamma$ rays can also be used to discriminate between positrons and $\beta^-$ or proton recoils to reduce the impact of cosmogenic decays or fast neutrons, respectively.

\section{Construction and deployment of 288 kg module}
\label{sec:deployment}

\begin{wrapfigure}{R}{0.5\textwidth}
    \centering
    \includegraphics[width=0.45\textwidth]{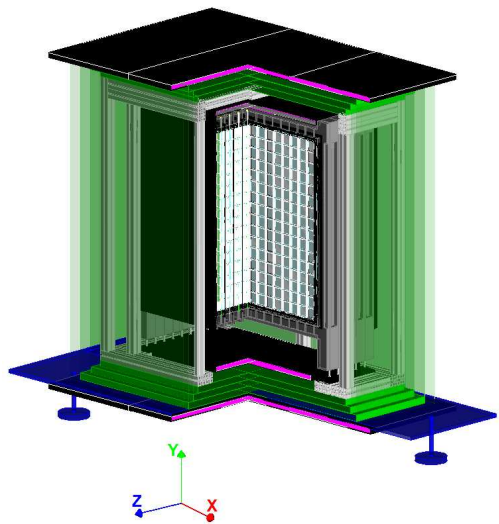}
    \caption{Diagram of the detector module showing 5 cm PVT cubes constructed inside aluminium frames and surrounded by HDPE shielding.}
    \label{fig:module}
\end{wrapfigure}

The 288 kg module was constructed in 2014, as explained in~\cite{celineEPS}.
The detector is constructed from 2304 PVT cubes, each coupled to a $^6$LiF:ZnS(Ag) screen and wrapped in a layer of Tyvek.
The PVT cubes are arranged in 9 detector frames each with $16\times16$ cubes, housed in an aluminium frame.
A 2D array of wavelength shifting optical fibres is inserted into grooves in the PVT cubes and coupled to SiPMs to detect the scintillation light.
During the construction phase, the mass of each individual PVT cube, lithium screen and Tyvek wrapper was precisely measured, allowing the number of protons in the detector to be calculated to a 1\% precision.

The signal from the SiPMs is amplified and digitised with a 14 bit resolution at a sampling rate of 65 MHz using custom electronics.
The digitised data is processed by a Gigabit Link Interface Board (GLIB)~\cite{GLIB}, which performs a per channel threshold based trigger, requiring coincident triggers from both a vertical and horizontal fibre within the same plane.
The triggered data is communicated to the data acquisition computing using the IPbus v2 protocol~\cite{ipbus}.

The nine detector frames are oriented vertically, mechanically joined and enclosed within a 9 cm thick HDPE neutron shield, as shown in \cref{fig:module}.
In addition muon veto scintillator panels are used as active shielding, with 4 panels installed above and 4 panels below the detector.
The detector was constructed off-site, and deployed at the BR2 reactor at the end of November 2014.
A period of detector commissioning followed the deployment.
The gain of each channel was measured and used to equalise the over-voltage of the 304 SiPMs.
The coincidence threshold trigger level was optimised to minimise the threshold whilst minimising the dead-time of the DAQ system due to high instantaneous trigger rates.

Data were taken for ten days in Feb 2015 with the reactor running at approximately 70 MW.
The BR2 reactor then shut down for an extended maintenance period that will last until 2016.
Following the reactor shutdown the detector took data in the same conditions as with the reactor on for a following two months.
This reactor off period can be used for estimating the main background event rates that are not related to the reactor running, mostly due to cosmic rays.
A radioactive source campaign was also performed.
Data were taken with AmBe and $^{60}$Co sources located around the detector.
The detector was then moved away from the reactor wall.
A PVT cube in the centre of the reactor was replaced with one containing an embedded $^{252}$Cf source.
The data from the source campaigns will supplement the reactor on/off data in estimating the neutron detection efficiency and the detector's energy scale and resolution.
The final source calibration runs were completed in September 2015.
Due to network bandwidth constraints, only a fraction of the data was available off site during running, transmitted by a combination of network and physical disk transfers.
Following the final calibration runs the detector and disk server holding the full data set were removed from the BR2 reactor building, making the full data set available for analysis for the first time.
Analysis of the full data sets is ongoing, but initial results on detector performance are reported below.

\section{Initial results with the 288 kg module}
\label{sec:initialresults}

\begin{wrapfigure}{R}{0.5\textwidth}
    \centering
    \includegraphics[width=0.45\textwidth]{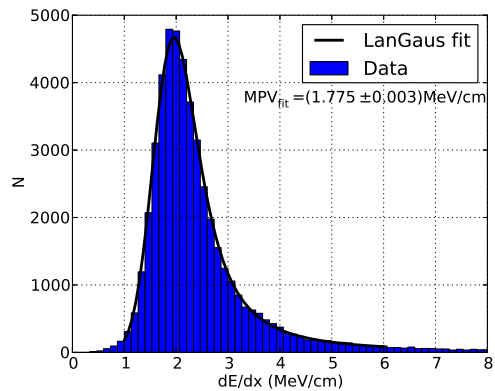}
    \caption{Distributions of measured $dE/dx$ from cosmic ray muons, compared with simulations.}
    \label{fig:muondEdx}
\end{wrapfigure}
Cosmic ray muons are a source of both background and calibration events for the SoLid detector.
Due to the high level of segmentation, cosmic ray muons passing through the detector deposit energy in many cubes along a straight track.
The energy deposited within a 5 cm cube due to a crossing muon is in the range $0 < E_{\mu} < 15$ MeV, covering the energy range of interest for the IBD positrons.
The crossing muons are tracked through the detector and their path length through a given cube calculated.
The measured signal amplitude and path length are used to construct a $dE/dx$ distribution for each PVT cube, shown in \cref{fig:muondEdx}.
These distributions have been used to perform an energy calibration across the detector, which is described in~\cite{danEPS}.
From this analysis a measurement of the detector's light yield can be extracted.
An average light yield per fibre of $13.0\pm 0.1$ photon/MeV has been found.


\begin{figure}[ht]
    \centering
    \includegraphics[width=0.65\textwidth]{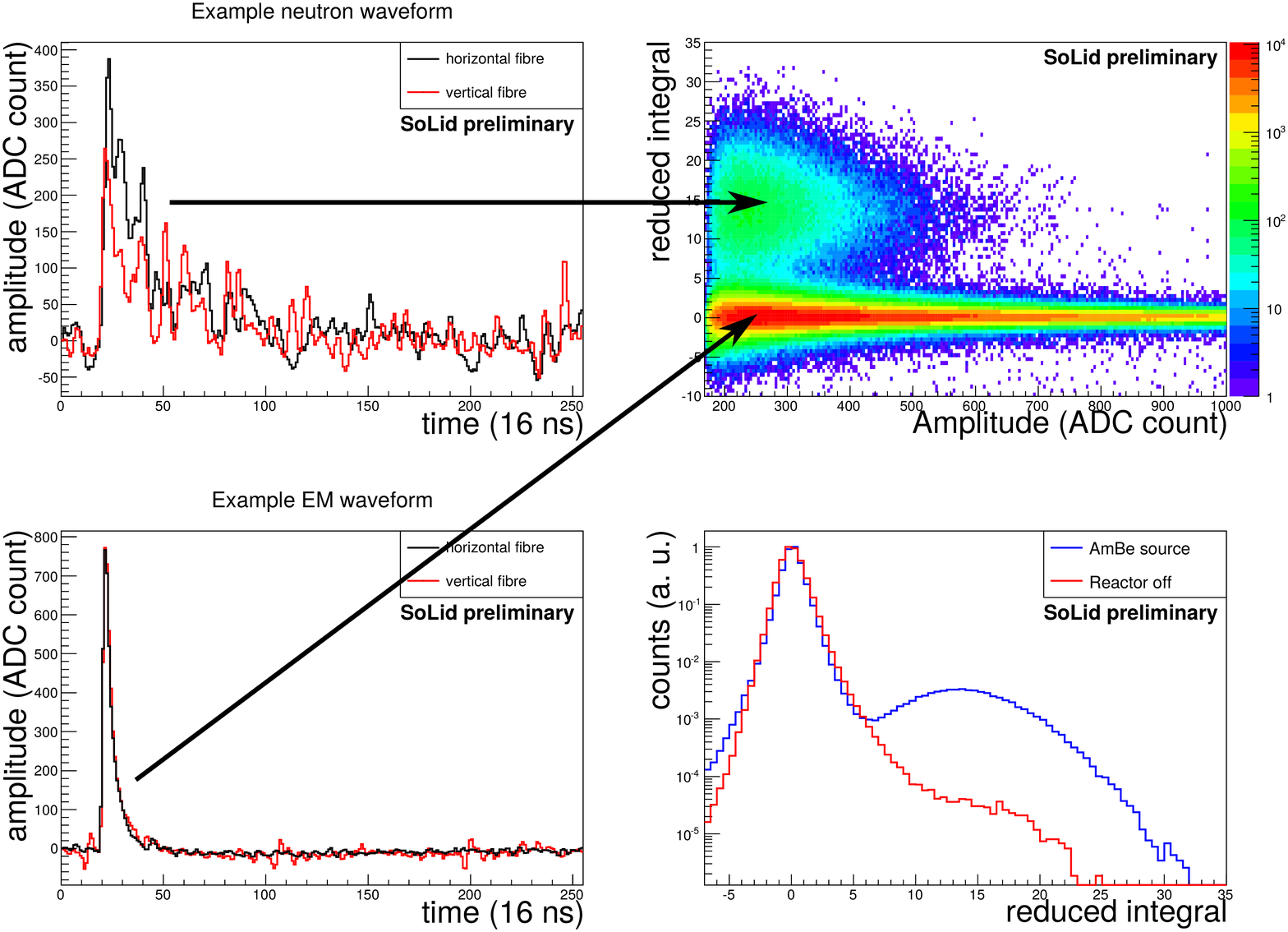}
    \caption{Neutron identification method, identifying waveforms with a large integral following correction for a standard PVT pulse shape.}
    \label{fig:neutronid}
\end{figure}

A neutron identification method has been developed using the reactor on/off and AmBe source data.
The method integrates a time window starting from the rising edge of an identified peak.
The integral is reduced by an amount modelled on a single PVT pulse of the same amplitude.
The result is therefore an integral of the light signal in the ZnS's slow decay tail, allowing neutron signals to be identified as those with a large value of the reduced integral.
In \cref{fig:neutronid} example ZnS and PVT waveforms are shown.
The neutron signals are seen in the large integral population in the integral versus amplitude histogram.
On the lower right a projection through this data set is shown, comparing data collected with the reactor off (with few neutrons) and with a neutron source deployed close to the detector.
In the latter case many more high integral waveforms are identified as neutrons.
This method and others are explained in more detail in~\cite{simonEPS}.

\begin{wrapfigure}{r}{0.5\textwidth}
    \centering
    \includegraphics[width=0.5\textwidth]{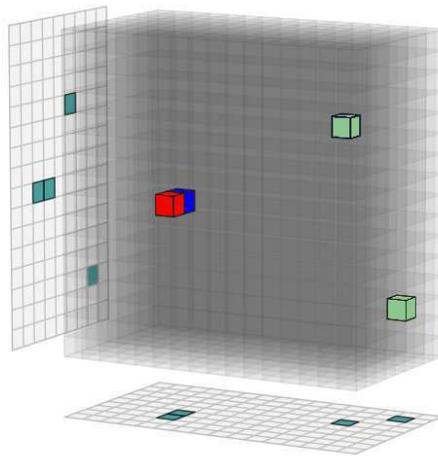}
    \caption{One of the first IBD candidate from reactor on data, with a positron candidate (blue) identified in close space-time proximity to a neutron signal (red).}
    \label{fig:ibdcandidate}
\end{wrapfigure}

IBD candidate events can be identified by the combination of a positron candidate and a neutron signal.
Initial IBD candidate events have been identified by requiring the positron candidate to be within a 10 cm radius from the neutron and to proceed the neutron signal by up to 250 $\mu$s.
One of the first IBD candidates is shown in \cref{fig:ibdcandidate}, where the neutron (red) and positron candidate (blue) cubes are identified, along with accidental $\gamma$ signals that occur within the 250 $\mu$s window, but too far from the neutron to be from an IBD event.
Due to the trigger threshold used it is not expected that any Compton scattering of the annihilation $\gamma$ rays will be seen in the data taken in 2015.

A full IBD analysis of this data using the reactor on data is in progress.
The reactor off data is being used for the estimation of cosmogenic and other background contributions to the IBD candidate collection.
The $\gamma$ and neutron source campaigns are also being used to confirm the energy scale calibration performed with the cosmic muon sample and to estimate the neutron detection efficiency.

\FloatBarrier

\section{Future plans for the SoLid experiment}
\label{sec:futureplans}


The full scale SoLid experiment is planned to start taking data in the second half of 2016, following the refurbishment of the BR2 reactor.
The experiment will be performed in two stages.
First an initial search will be performed using a two tonne detector based upon the 288 kg module deployed this year.
One year of data (roughly half reactor on, half off) will be used for the initial search.
Then an additional tonne of detector mass with an improved energy resolution will be deployed.
The full 3 tonne detector will take data for three years, providing a world-leading sensitivity for anti-neutrino oscillations.

\begin{wrapfigure}{R}{0.5\textwidth}
    \centering
    \includegraphics[width=0.45\textwidth]{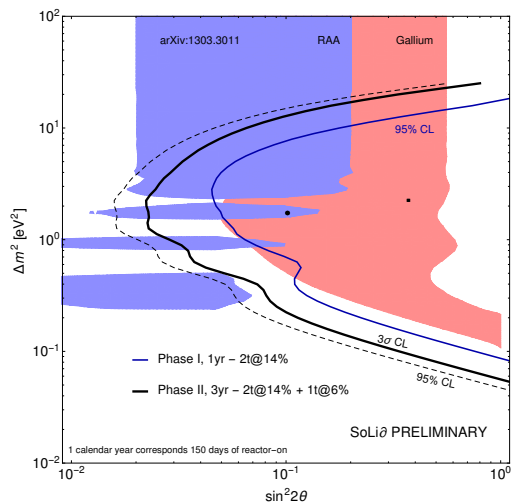}
    \caption{The estimated sensitivity that the SoLid experiment will have to 3+1 sterile neutrino models, as a function of the oscillation parameters $\sin^2(2\theta)$ and $\Delta m^2$. The 95\% confidence level after the initial search and the $3\sigma$ sensitivity after the precision search are shown. These can be compared to the 95\% confidence level parameter space from the reactor and gallium anomalies shown in blue and red, respectively.}
    \label{fig:sensitivity}
\end{wrapfigure}

The initial detector modules will be a larger version of the 288 kg module deployed this year, with an upgraded read-out system capable of triggering a time window read-out based on an FPGA-level neutron identification.
The number of WLS fibres will be doubled so that the positron energy will be reconstructed from the sum of the signals from four fibres.
The expected total light yield should be 50 photon / MeV.
The statistical contribution to the energy resolution will therefore be 14\% at 1 MeV.
The positron has a high probability to be contained within a single PVT cube and a negligible fraction of the energy carried by the 511 keV annihilation $\gamma$-rays is also deposited within the cube.
The photon statistics are therefore expected to be the dominant factor in the positron (and therefore neutrino) energy resolution.
With one year of data taking (150 days with the reactor running at 70 MW) and initial oscillation search will be performed.

During the initial search's data taking phase an additional tonne of detector, designed to have a 6\% energy resolution at 1 MeV, will be constructed.
The high resolution detector will be deployed at the distance of closest approach (roughly 5.5 m) from the reactor core, with the 2 tonne original detector relocated to be 6.5 - 10 m from the reactor.
A precision oscillation search will be performed with three years (450 days with the reactor running at 70 MW) worth of data.

The expected sensitivity to oscillations has been calculated and is shown in \cref{fig:sensitivity}.
The calculation assumes a 40\% IBD detection efficiency and a signal to noise ratio of 3.
The background signal is an equal contribution of a $1/E_{\bar{\nu}}^2$ shape and one flat in anti-neutrino energy.
A 2\% relative normalisation error is assumed in the calculation using both detectors.
A 95\% confidence level contour for the initial search, and a 3$\sigma$ significance contour for the precision search are shown in \cref{fig:sensitivity}.
Also shown are the 95\% CL allowed region for the reactor and gallium anomalies, along with their best fit values.

The initial search will rapidly provide 95\% CL coverage of a large proportion of the anomaly allowed regions.
The precision search will cover the anomaly region with 3 $\sigma$ significance.
The precision experiment uses an optimised detector.
The improved energy resolution of the additional detector module, deployed close to the reactor core extends the sensitivity to large $\Delta m^2$ values.
The large detector mass and 5.5 m to 10 m range of baselines provided by the three tonne detector gives sensitivity down to low $\sin^2(2\theta)$ values.
The result of the optimisation is that the three tonne detector will provide a world-leading sensitivity to oscillations despite a modest overall cost.

\section{Conclusions}

The SoLid Collaboration is building a novel, highly segmented, composite scintillator detector in order to resolve the reactor neutrino anomaly.
A 288 kg detector module have been constructed and run at the BR2 research reactor at SCK$\bullet$CEN in Belgium.
In 2015 the 288 kg module took data with the reactor on, the reactor off and with a number of radioactive sources.
This data is being used for an ongoing anti-neutrino analysis.
Initial results from identifying neutron waveforms and using muon tracks for calibration were presented.
In 2016 the initial modules of a three tonne detector will be deployed at the BR2 reactor.
Running with the full detector for three years will allow the collaboration to make a world-leading, highly sensitive search for short baseline oscillations, allowing the collaboration to resolve the reactor neutrino anomaly

\section{Acknowledgements}

The SoLid Collaboration would like to thank SCK$\bullet$CEN and the BR2 reactor support staff for the very successful operation of the BR2 reactor at which the SoLid detector is located.
The SoLid collaboration would like to acknowledge the funding received from the Carnot Institutes, the Flemish Hercules Foundation (Vlaamse Herculesstichting), the Research Foundation - Flanders (FWO) and the Science and Technology Facilities Council, and the University of Oxford's particle physics sub-department.
In addition the author would like to acknowledge funding received from Merton College, Oxford.

%


\end{document}